# High-dimensional acousto-optoelectric correlation spectroscopy reveals coupled carrier dynamics in polytypic nanowires

Maximilian M. Sonner[1], Daniel Rudolph[2], Gregor Koblmüller[2], Hubert J. Krenner[1,3]

[1]Lehrstuhl für Experimentalphysik 1, Institut für Physik, Universität Augsburg, Universitätsstraße 1, 86159 Augsburg, Germany.

[2]Walter Schottky Institut and Physik Department, Technische Universität München, Am Coulombwall 4, 85748 Garching, Germany.

[3]Physikalisches Institut, Westfälische Wilhelms-Universität Münster, Wilhelm-Klemm-Straße 10, 48149 Münster, Germany

## Abstract
The authors combine acousto-optoelectric and multi-channel photon correlation spectroscopy to probe spatio-temporal carrier dynamics induced by a piezoelectric surface acoustic wave (SAW). The technique is implemented by combining phase-locked optical micro-photoluminescence spectroscopy and simultaneous three-channel time resolved detection. From the recorded time correlated single photon counting data the time transients of individual channels and the second and third order correlation functions are obtained with sub-nanosecond resolution. The method is validated by probing the correlations SAW-driven carrier dynamics between three decay channels of a single polytypic semiconductor nanowire on a conventional LiNbO$_3$ SAW delay line chip. The method can be readily applied to other types of nanosystems and probe SAW-regulated charge state preparation in quantum dots, charge transfer processes in van der Waals heterostructures or other types of hybrid nanoarchitectures.

## Main Text

### 1 Introduction
Surface acoustic waves (SAWs) are a versatile tool to probe and manipulate the physical properties of a wide variety of systems [1]. For instance, these surface confined acoustic waves have found widespread application for the study of semiconductor nanosystems [2–4] at radio frequencies. When these waves propagate on a piezoelectric, strain induces a gyrating electric field [5]. This electric field in turn efficiently ionizes excitons [6] and induces spatiotemporal dynamics of such dissociated electrons and holes in semiconductor nanostructures [7]. The induced dynamics strongly modulate the optical emission in the time domain and their characteristic fingerprints can be detected in photoluminescence (PL) experiments [7–9]. In acousto-optoelectric spectroscopy (AOES), the underlying spatiotemporal dynamics are compared to calculation which allows to determine the transport mobilities of electrons and holes inside the studied nanostructure [10,11]. To date, this versatile method is limited to systems with a single transport channel which can be selected via its characteristic emission signal in the optical spectrum. In many systems, different transport and/or recombination channels coexist. One prominent example are epitaxial semiconductor nanowires (NWs). In this system, crystal phase mixing (polytypism) gives rise to the formation of barriers and carrier localization [12–14]. In NW-based core-multishell heterostructures, band edge energies are strongly affected by unavoidable size and compositional fluctuations [15–19]. Employing AOES, first studies have been conducted [10,17] which were restricted to individual recombination channels. To unravel coupled dynamics, this limitation can be overcome by high-order correlation techniques, which are widely applied in optics to study a broad variety of systems [20–24].

In this article, we advance the AOES to a multi-channel platform to monitor the time-dependent SAW-modulated emission of up to three spectrally isolated decay channels. Using a polytypic GaAs-(Al)GaAs core-shell NW as a model system, we are able to directly and simultaneously observe coupled carrier dynamics. Moreover, we show that we can directly obtain high-order correlations for these data revealing clear fingerprints of the underlying coupled dynamics.

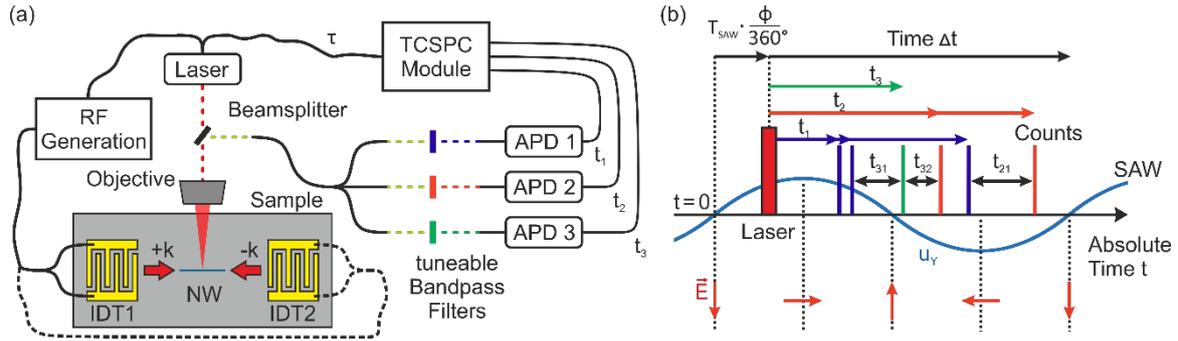

**Figure 1:** (a) *Schematic of the 3CXC AOES* – Phase-locked, micro-photoluminescence spectroscopy probes the nanosystem (here a single NW). Emission of different recombination channels are spectrally isolated and detected by APDs and recorded by multi-channel TCSPC electronics. (b) *Timing scheme* – The local phase of the SAW oscillation (blue) sets absolute time, $t$, of the experiment. Laser excitation (red bar) is delayed by $T_{SAW} \cdot \phi/360°$ and individual time stamps of subsequent detection events on each channel are registered relative to the laser excitation.

## 2 Experimental implementation

Our experimental setup for three-channel cross correlation (3CXC) acousto-optoelectric spectroscopy (AOES) is depicted in Figure 1(a). Propagating SAWs are generated on a SAW-chip by applying a resonant radio frequency (rf) voltage to interdigital transducers (IDTs). As shown in Figure 1(a), we typically use two nominally identical IDTs forming a delay line. This configuration allows for the generation of two SAW beams which propagate in opposite directions. In the following we will refer to these two directions as $+k$ (left to right) and $-k$ (right to left) propagating SAWs. In this work, we use a SAW delay line which was fabricated on a commonly used strong piezoelectric substrate, YZ-cut LiNbO$_3$ substrate (phase velocity $c_{SAW} \simeq 3500 \frac{m}{s}$). IDTs are designed for the excitation of SAWs with a wavelength of $\lambda_{SAW} = 18$ µm and a corresponding acoustic period (frequency) of $T_{SAW} = 5.08$ ns ($f_{SAW} = 197$ MHz). The system studied is transferred onto the active region of the SAW-chip in between the two IDTs. In this work, we use a polytypic GaAs-(Al)GaAs core-shell NW which is characterized in Figure 2. The SAW chip is mounted in a low temperature ($T = 8$K) micro-photoluminescence setup, electron-hole pairs are photogenerated using an electrical triggered pulsed diode laser pulse width $\tau_{laser} = 90$ ps, wavelength of $\lambda = 661$ nm) focused to a diffraction limited spot of about ~1.5 µm [25,26]. A key characteristic of AOES is that charge carriers can be photogenerated at a well-defined local phase of the SAW, i.e. time during the SAW's oscillation. This phase-locking is achieved by referencing the frequency of the SAW to a multiple integer of the repetition frequency of the laser, $n \cdot f_{Laser} = f_{SAW}$, with $n$ being integer. Similar phase locked excitation can be also realized using free-running mode-locked lasers with constant repetition rate [27,28]. The emission of the studied nanostructure is collected by the same objective and split via a symmetric 1x3 fiber beam splitter. Each of its three outputs is spectrally filtered by a tunable bandpass filter (FWHM bandwidth of 3.0 nm) to isolate emission of individual emission bands. These signals are detected by single-photon silicon avalanche photodetectors (APDs) with a time resolution $< 350$ ps. Time-correlated single photon counting (TCSPC) is performed by a five-channel time tagger. As stated above, the (nano)system studied can be optically excited at a well-defined time during the acoustic cycle. The full timing scheme of our experiment is detailed in Figure 1(b). The absolute time, $t$, is set by the SAW at the position of the laser spot on the system studied. Here, we use the vertical displacement component $u_Y$ [blue line in Figure 1(b)] as reference. As the absolute time, $t$, progresses the vertical displacement oscillates $u_Y \propto \sin 2\pi f_{SAW} t$ and the electric field vector $\vec{E}$ [red arrows in Figure 1(b)] gyrates counterclockwise close to the surface for the LiNbO$_3$ substrate used in this work. The phase, $\phi$, between the laser and the SAW at the position of the laser spot focused onto the system studied can be set to any value between $\phi = [0°, 360°]$. The time stamps of all detection events ($t_i$) registered by one of the detectors (APD $i$) are recorded individually as shown in Figure 1(b). Thus, we can perform a full analysis of the detection events on the three channels. First, we can perform conventional single channel analysis on channel $i$ by referencing the registered events to the laser excitation. Most importantly, we can determine rigorously correlations between the three channels. This is possible, because we do not have to perform recordings for individual, single channels but all signals are recorded *simultaneously* for all channels in a single run. In our experiment analyzing three channels, we determine time differences (relative to the laser pulse) $t_{ij}$ between detection events of channels $i$ and $j$ as shown in Figure 1(b). Then, we use these time differences to calculate three second order intensity correlations

$$g_{ij}^{(2)}(t_{ij}) = \frac{\langle I_i(t) I_j(t+t_{ij}) \rangle}{\langle I_i(t) \rangle \langle I_j(t) \rangle} \quad \text{(Equation 1)}$$

and the third order intensity correlations

$$g^{(3)}_{1,2,3}(t_{21}, t_{31}) = \frac{\langle I_1(t)I_2(t+t_{21})I_3(t+t_{31})\rangle}{\langle I_1(t)\rangle\langle I_2(t)\rangle\langle I_3(t)\rangle} \qquad \text{(Equation 2)}.$$

## 3   Nanowire model system

In this article, we focus on the SAW-driven carrier dynamics in a polytypic NW with different recombination channels as grown by molecular beam epitaxy. The polytypism is characterized by intermixing of zincblende (ZB) and wurtzite (WZ) crystal phase segments of random length along the NW axis, which leads to a static type-II modulation of the conduction (CB) and valence band (VB) edges. This type-II band edge modulation is known to open different recombination channels which can be distinguished from their optical transition energies [12,29–33] and to leave characteristic fingerprints in the AOES [10]. To validate our method, we transfer GaAs/(Al)GaAs core/shell NWs onto our SAW-chip. For our experiments, we select a NW located in between the two IDTs and which is aligned parallel to the propagation of the SAW to induce acoustoelectric transport [34].

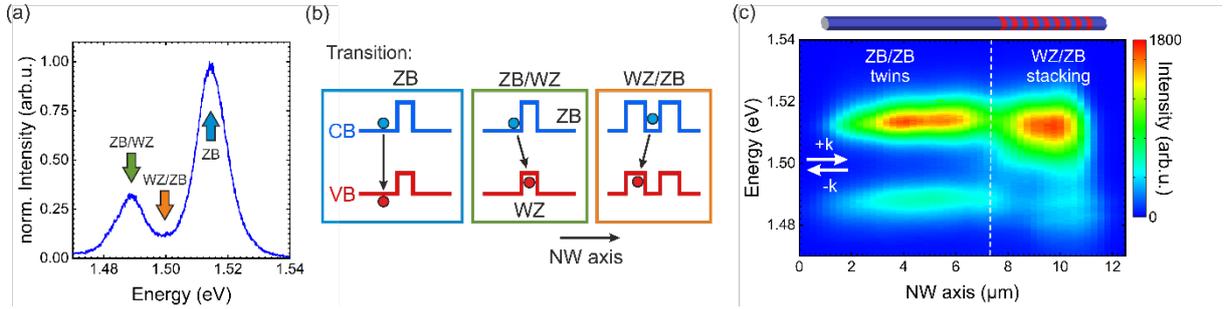

**Figure 2:** *Characterization of the studied NW* – (a) Photoluminescence spectrum observed at the center of the studied polytypic GaAs/AlGaAs core/shell NW. (b) Schematic of the bandstructure along the axis of a NW core and recombination processes corresponding to the three different recombination bands. (c) Spatial PL map along the NW. The white dashed line marks transitions for ZB with occasional twins to a mixed WZ/ZB phase as shown by the schematic. $+k$ and $-k$ propagating SAWs are parallel aligned to the NW.

A µ-PL spectrum recorded at low optical pump power densities ($2.8\,\text{W}/\text{cm}^2$) at the center of the NW is depicted in Figure 2(a) and shows an emission band typical for polytypic GaAs NWs [12,30–33]. The type-II band alignment along the NW axis results in the localization of electrons and holes in ZB and WZ regions, respectively. In the presence or absence of quantum confinement, three dominant recombination channels exist, which are shown schematically in Figure 2(b): (i) Recombination of electrons and holes in extended ZB-phase regions [left panel, blue]. This process can be attributed to the strong PL peak at 1.514 eV marked by a blue arrow in the spectra in Figure 2(a)]. (ii) Spatially indirect recombination between quasi-free electrons in extended ZB-phase regions and quantum confined holes in thin WZ-segments [center panel in Figure 2 (b), green]. This recombination contributes significantly in predominantly ZB-phase regions containing several twin defects which can be understood as a monolayer thick insertion of the WZ material in a ZB matrix [14] or short WZ segments. We refer to this process as ZB/WZ recombination in the following. Emission from this ZB/WZ recombination is observed at lower energy at $E_{ZB/WZ} = 1.488$ eV compared to the ZB-transition and is marked by a green arrow in Figure 2(a). (iii) Spatially indirect recombination between quantum confined electrons in short ZB-regions and quantum confined holes in short WZ-regions [right panel in Figure 2(b), orange] occurs in highly mixed regions of a NW. We refer to this process as WZ/ZB in the following. In contrast to indirect exciton transitions in ZB/WZ regions discussed before, in the mixed WZ/ZB region both carrier species are confined. This results in a shift to higher (lower) energies compared to the ZB/WZ (ZB) emission peak. Thus, the emission energy is expected in between that of the ZB- and ZB/WZ bands, as marked by the orange arrow in Figure 2(a).

Next, we analyze the evolution of the µ-PL spectra recorded along the axis of the NW in Figure 2(c). The emission intensity is color-coded and plotted as a function of photon energy (vertical axis) and position along the NW axis (horizontal axis). In the left and the center part of the NW, PL emission peaks at 1.514 eV and 1.488 eV are observed. Thus, we conclude that in these parts of the NW the ZB crystal structure dominates (ZB peak) and a few occasional rotational twins occur (ZB/WZ peak). Towards the right part, a transition in the crystal structure occurs from predominantly ZB to a highly mixed phase containing extended WZ segments embedded in the ZB crystal. This results in a pronounced change in the PL spectrum at the position highlighted by the white dashed line. Here, the PL emission slightly shifts from the pure ZB peak observed on the left towards lower energies and broadens. This new emission peak is attributed to the indirect exciton transitions in the mixed WZ/ZB region. Moreover, for thin segments motional quantization contributes and the energy of electrons and holes shift away from their respective band edges. This in turn leads to a shift of the emission energy. This clear transition from near pristine ZB to mixed ZB/WZ phases, confirmed by the PL data, is shown in the schematic in the lower part of Figure 2 (c). These PL characterization data clearly confirms that the three characteristic emission bands corresponding to different recombination channels can be spectrally filtered. Moreover, the selected NW shows a clear transition from a near-pristine ZB crystal structure to a highly mixed WZ/ZB phase. These properties make this NW particularly well suited to validate 3CXC AOES.

## 4 Multi-channel spectroscopy

### 4.1 Single channel detection

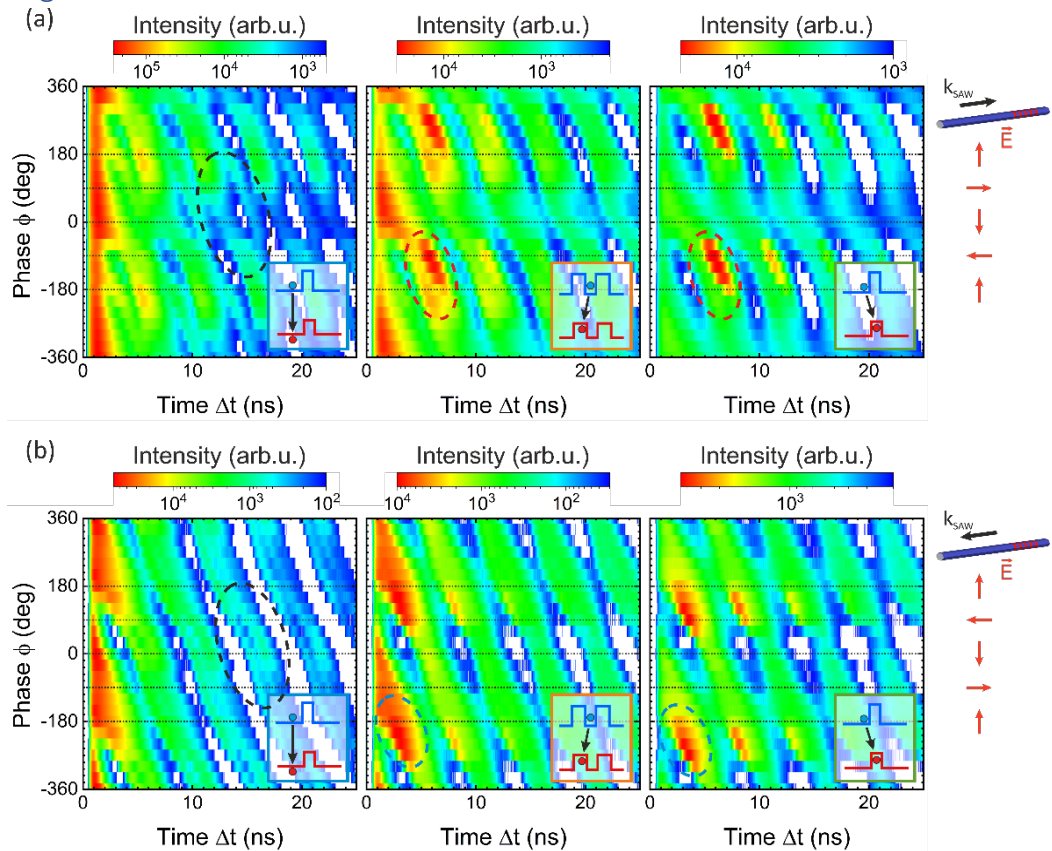

**Figure 3:** *Single channel AOES* – Stroboscopic PL spectra of the ZB (left panels), the indirect ZB/WZ (center panels) and indirect WZ/ZB (right panels) exciton transitions at the center of the NW for (a) $+k$- propagating SAW and (b) $-k$- propagating SAW and phases of $\phi = [-360°, 360°]$. The electric field orientation at photoexcitation is shown for different phases $\phi$ on the right. The recorded transients show SAW induced oscillations and exhibit clear dependences on the SAW propagation directions (marked by black dashed ellipses and black and red arrows).

Next, we validate our technique by applying $+k$- and $-k$-propagating SAWs. The orientations of these SAWs with respect to the NW axis are shown in Figure 2(c). We begin by analyzing the SAW-induced emission dynamics for each of the three emission bands individually at the center of the NW. A full set of AOES data of the ZB transition (left panels), the WZ/ZB emission (center panels) and the ZB/WZ transition (right panels) is presented in Figure 3. These different recombination channels are shown as insets. Figures 3 (a) and (b) show data for $+k$- and $-k$-propagating SAWs, respectively. The recorded intensity (normalized to corresponding maximal intensity) is plotted in false-color representation as a function of the phase $\phi$ (vertical axis) and the time after photoexcitation $\Delta t$ (horizontal axis). On the right, the electric field orientation is shown for different $\phi$ marked by the horizontal dotted black lines in the data panels. As we tune the phase over two full acoustic cycles $\phi = [-360°, +360°]$, we resolve clear SAW induced intensity modulations for all three transitions and both SAW directions. These modulations are characteristic fingerprints of transport of electrons and holes along the NW axis which are driven by the oscillating electric field parallel to the NW axis [7,10,34]. Interestingly, the data shows that these characteristic features exhibit both marked similarities and differences for the three recombination channels and SAW propagation directions. First, the intensity modulations of the same recombination channel clearly change when the phase, $\phi$, i.e. the time during the SAW cycle the laser creates electron-hole pairs, is tuned. This observation nicely demonstrates that the orientation of electric field shown on the right at the position of the laser spot at the time of the laser excitation programs the subsequent SAW-driven carrier dynamics [8,9]. Second, the intensity modulations of the same recombination channel depend on the propagation direction of the SAW. One prominent example for the ZB transition is marked by the dashed black ellipses in Figure 3(a) and (b) showing a $T_{SAW}$-periodic intensity modulation with two recombination events per SAW cycle for both propagation directions. The temporal delay of both events clearly varies as the phase, $\phi$ is tuned. Yet, this particular feature is significantly more pronounced in the data of the $+k$-propagating compared to those of the $-k$-propagating SAW. These findings can be qualitatively understood by considering the morphology of the NW. The $+k$-propagating ($-k$-propagating) SAW effectively induces transport toward (away from) the highly mixed region and the holes located within and near the region in the $+x$-direction. Since the type-II bandedge modulation creates a barrier for electrons in the CB, the transport for electrons is reduced along the NW for $+k$-propagating SAWs compared to the $-k$ propagating SAW. For the latter, electrons are transported along the NW away from the quasi-stationary holes reaching the lower end of the NW and recombine non-radiatively at the open surface. This reduces the total overlap time of electrons and holes. Another clear directional dependency is observed for the indirect transitions. For example, we observe distinct, directional features in the ZB/WZ and

WZ/ZB data, marked by dashed red and blue ellipses. These signals are more pronounced for $+k$- propagating SAW than for $-k$- propagating SAW and result from the charge carriers being transported toward the highly mixed region for $+k$- propagating SAWs. Furthermore, time-delayed emission of both indirect transitions occurs when the electric field points away from the highly mixed region at the time of photoexcitation on the right (marked by dashed red and blue ellipses). At these times electrons are accelerated towards the barriers and retain inside the NW. The sense of the electric field's gyration gives rise to the observed shift in $\Delta t$ at which the signal is detected.

## 4.2 Two-channel correlation analysis

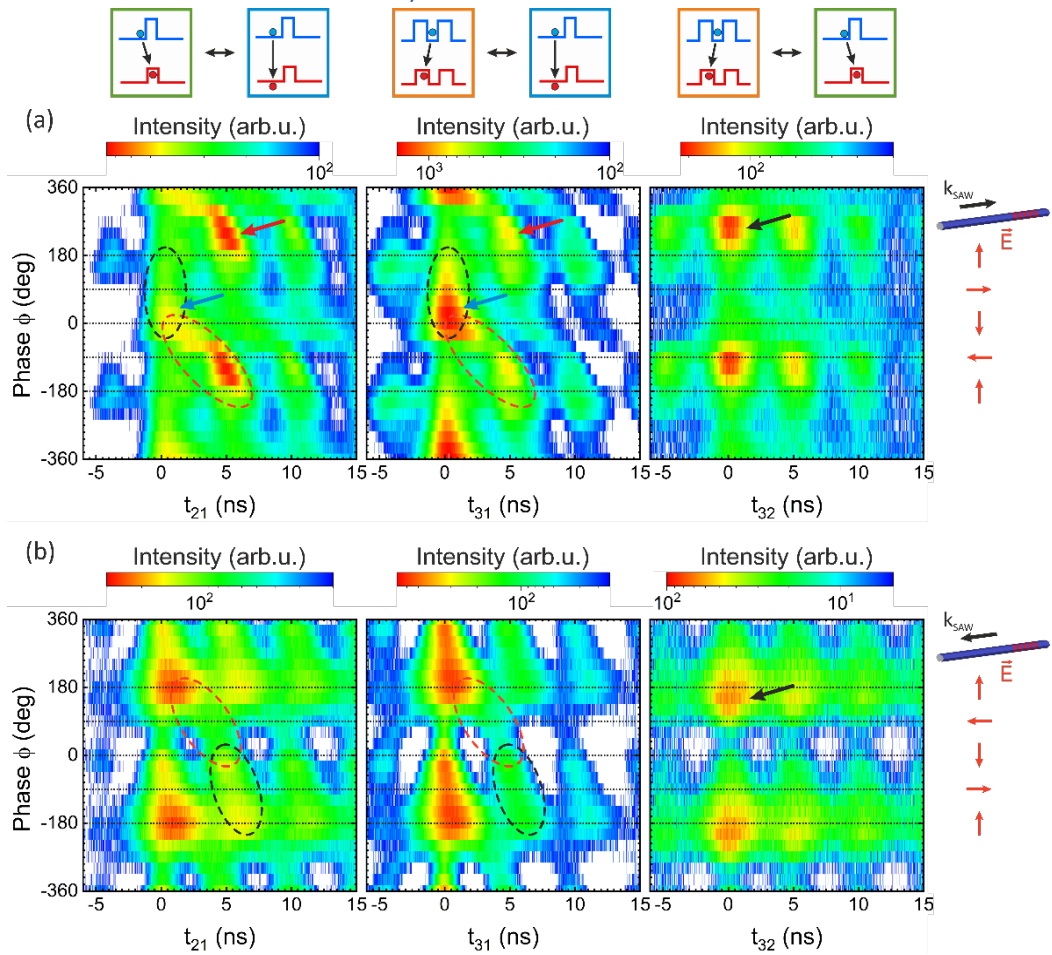

**Figure 4:** *2-channel correlation spectroscopy* – $g_{ij}^{(2)}(t_{ij})$ of the indirect ZB/WZ exciton - ZB exciton ($i,j = 2,1$, left panels), the indirect WZ/ZB exciton - ZB exciton ($i,j = 3,1$, center panels) and indirect WZ/ZB exciton - ZB/WZ exciton ($i,j = 3,2$, right panels) as a function of the phase $\phi$ for (a) $+k$- and (b) $-k$- SAW propagation direction. The electric field orientation at photoexcitation is shown on the right. Characteristic correlation features are highlighted by arrows and ellipses.

In the next step, we investigate the second order intensity correlations between the three transitions of the dynamically driven NW system. To this end, we calculate the second order correlation functions $g_{ij}^{(2)}(t_{ij})$ as defined by (Equation 1), between events of two transitions $i,j$ for both SAW propagation directions. $i = 1, 2, 3$ correspond to the ZB, ZB/WZ and WZ/ZB transitions, respectively. Here, $t_{21}$, $t_{31}$ and $t_{32}$ are the time differences between the indirect ZB/WZ–ZB, WZ/ZB–ZB and WZ/ZB–ZB/WZ transitions. In Figure 4 (a) and (b), we plot the recorded $g_{ij}^{(2)}(t_{ij})$ (normalized to corresponding maximal intensity) of the SAW-driven NW as a function of $\phi$ in false-color representation for $+k$- and $-k$- SAW propagation direction, respectively. Again, the orientation of $\vec{E}$ at the time of photoexcitation with respect to the NW axis is shown on the right. As the phase is tuned, clear SAW induced $T_{SAW}$-periodic and phase dependent modulations with characteristic features are observed for both propagation directions in all $g_{ij}^{(2)}(t_{ij})$ data. These modulations in all $g_{ij}^{(2)}(t_{ij})$ data can be understood by considering the initial distributions of electrons and holes at the time of photoexcitation. We start our evaluation for the SAW $+k$-propagating towards the stacking faults. Obviously, the measured correlations between the two indirect transitions $g_{32}^{(2)}$ (ZB/WZ–WZ/ZB, right panels) significantly differ from those obtained either by the indirect transitions or the ZB emission $g_{i1}^{(2)}$ (left and center panels). $g_{32}^{(2)}(t_{32})$ exhibits clear $T_{SAW}$-periodic oscillations, $g_{32}^{(2)}(t_{32} = n \cdot T_{SAW})$, with $n$ being integer. These oscillations prove that the two indirect transitions are coupled and occur at the same time or at integer multiples of $T_{SAW}$ for each SAW phase. This can be understood considering that for both types of indirect transitions holes are captured within single twin plane defects or WZ segments. Both localization sites are predominantly located in the same region of the NW. Electrons in contrast are accelerated along the NW by the electric field and periodically shuffled back and forth to the region with high rotational twin density (ZB/WZ) and the highly mixed WZ/ZB region. As a result, holes recombine almost simultaneously with the oscillating electrons giving rise to the $T_{SAW}$-periodic oscillations of $g_{32}^{(2)}(t_{32})$. Furthermore, the $T_{SAW}$-periodic oscillations exhibit a maximum intensity at $\phi = 270°$ (marked by black arrow). At this phase, electrons are accelerated at the time of photoexcitation towards the highly mixed region and are captured within both types of stacking faults. This efficient capture probability favors simultaneous recombination via both indirect transitions at the same time and, thus, an increase of $g_{32}^{(2)}(t_{32})$.

In contrast, both $g_{i1}^{(2)}$ data (left and center panels) show that either of the indirect transitions and the ZB emission is temporally delayed at distinct phases. As holes in the ZB phase are more mobile, the hole distribution within the pure ZB part of the NW is determined by the electric field orientation at photoexcitation. For phases ranging between $\phi = 0°$ and $180°$, holes are accelerated towards the region with a high stacking fault density on the right by the axial component of electric field. As discussed above, electrons are shuffled back and forth along the NW axis and recombine almost simultaneously with holes in the ZB part near the highly mixed WZ/ZB region or in the region with a high twin density (ZB/WZ), respectively. Hence, the correlations $g_{i1}^{(2)}$ (left and center panels) also show that each indirect and direct transition occur at the same time or at integer multiples of $T_{SAW}$, marked by black dashed ellipses. For phase between $\phi = 180°$ and $360°$, the electric field component parallel to the NW axis is pointing towards the ZB region with a maximum at $\phi = 270°$. This results in an acceleration of the photogenerated holes toward the ZB part of the NW. Consequently, holes in the ZB part are separated from holes trapped within stacking faults. For these phases, electrons oscillate back and forth between holes in the ZB part and holes within the region with high stacking fault density. The electron dynamics lead to a temporal delayed emission of the direct and both indirect transitions. The magnitude of the electric field component parallel to the NW axis decreases for phases $\phi > 270°$. Thus, the spatial separation of holes in the ZB part and holes trapped within the region with high stacking fault density reduces. Consequently, the temporal delay between the direct and both indirect transitions reduces as well, which manifests itself in a clear shift of the $g_{i1}^{(2)}$ peaks marked by red dashed ellipses. Moreover, both correlations $g_{i1}^{(2)}$ exhibit a high intensity at $\phi \approx 270°$ marked by red arrows. As discussed in the context of $g_{32}^{(2)}$, the electron density is increased within the stacking faults because at the time of photoexcitation the electric field oriented in the opposite direction. This in turn results in an increased recombination rate of both indirect transitions. In addition, we first find that $g_{31}^{(2)}(t_{31})$ is maximum for $\phi$ ranging between $0°$ and $180°$ and $t_{31} = 0$, (marked by blue arrows), In contrast, $g_{21}^{(2)}(t_{21})$ is maximum for $\phi$ ranging between $\phi = 180°$ and $360°$ and time differences of $0 < t_{21} < T_{SAW}$, marked by red arrows. Both observations can be well understood by considering the electric field component parallel to the NW axis: in the first case, holes are transferred along the NW from the region with high rotational twin density (ZB/WZ) to the highly mixed WZ/ZB region when the electric field is orientated towards the highly mixed region at the time of photoexcitation. Hence, the hole density with the stacking faults is increased in the highly mixed region (WZ/ZB transitions) and lowered in region with high rotational twin density (ZB/WZ transitions). In the second case, the electric field is orientated towards the region with high rotational twin density. This gives rise to an increase of the hole density at rotational twin defects while the hole density is reduced in the highly mixed region.

When reversing the SAW direction to $-k$- SAW propagation in Figure 4 (b), the $g^{(2)}_{32}(t_{32})$ data in the right panel are strikingly similar to those for the $+k$-propagating SAW in Figure 4 (a). This is expected because the underlying shuffling electron dynamics remain unaffected and, thus, do the resulting correlations. In contrast, the $g^{(2)}_{21}$ and $g^{(2)}_{31}$ correlations shown in the left and center panels are reversed because the sense of the gyration of the electric field vector is reversed (cf. schematics on the right). For the phase between $\phi = 180°$ and $360°$, the electric field points towards the stacking faults at the time of photoexcitation. Thus, the indirect exciton transitions and the free exciton transitions occur simultaneously (or at integer multiples of $T_{SAW}$) as marked by black dashed ellipses. In contrast, the delay between recombination events in this setting shifts from $T_{SAW}$ to 0 for phases from $\phi = 0°$ to $180°$ as marked by red dashed ellipses. This is in contrast to the $+k$-propagation, since here the axial electric field component pointing to the left accelerates electrons in the opposite direction toward the WZ/ZB region.

## 4.3 Three-channel correlation analysis

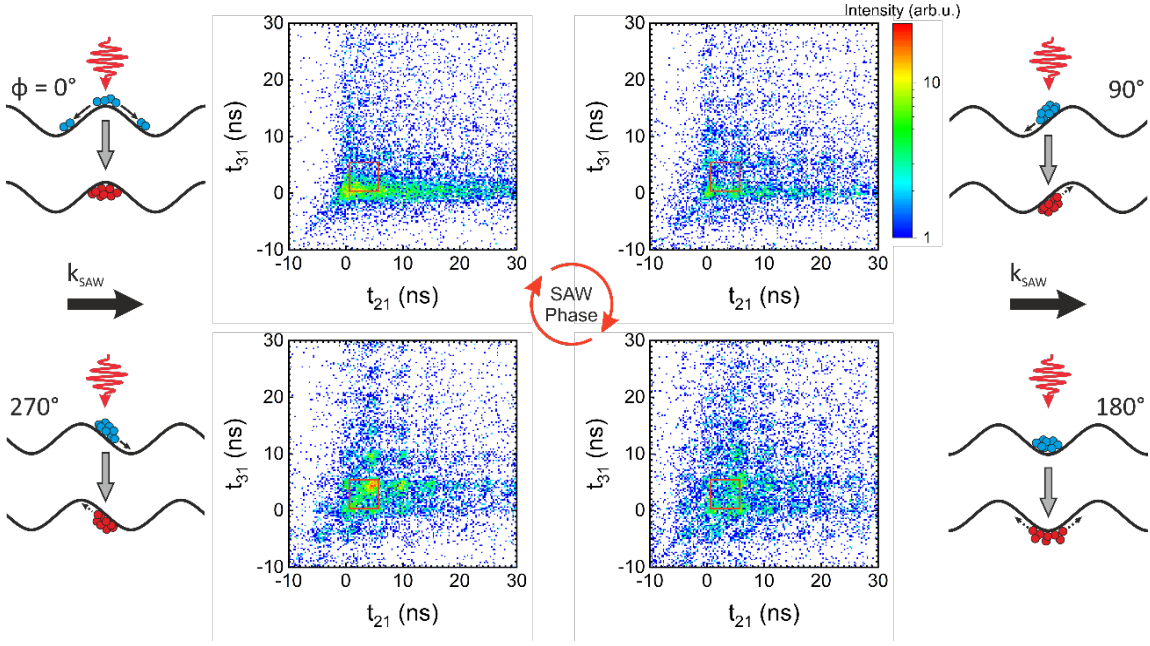

**Figure 5:** *3-channel correlation spectroscopy* – $g^{(3)}_{1,2,3}(t_{21}, t_{31})$ of all three recombination channels for four characteristic values of $\phi$ for a $+k$-propagating SAW (color scale applies to all panels). The band edge modulation by the SAW is shown as schematics. The red square marks $T_{SAW}$ setting the period of the observed oscillations.

Finally, we determine the third order intensity correlations $g^{(3)}_{1,2,3}(t_{21}, t_{31})$ of all three transitions as defined by (Equation 2). Here, $t_{21}$ and $t_{31}$ are the time differences between a detection of the first photon of the ZB transition and the detection of a photon of the indirect ZB/WZ and WZ/ZB recombination, respectively. In Figure 5 we plot the obtained $g^{(3)}_{1,2,3}$ color-coded as a function of $t_{21}$ and $t_{31}$ for four characteristic phases $\phi$ of a $+k$-propagating SAW. The respective excitation conditions are shown as schematics next to each data panel. All four data sets reveal $T_{SAW}$-periodic modulations in $t_{21}$ and $t_{31}$ marked by a red square. The observation of these modulations is a distinct and characteristic fingerprint of a SAW-induced charge carrier transport. Furthermore, the measured correlations strongly depend on $\phi$. At $\phi = 0$, holes are generated in a stable maximum of the valence band while electrons are generated in an unstable maximum of the conduction band. Thus, electrons rapidly redistribute toward their adjacent stable minima along the NW axis while the less mobile holes remain predominantly at the point of photogeneration [10]. Thus, the three-photon correlations predominantly occur at $t_{21} = t_{31} = 0$. For $t_{31} = 0$, the signal decays slowly for $t_{21} > 0$. This is consistent with the two-photon correlation data in Figure 4 (a) where a strong signal is observed for $g^{(2)}_{31}(t_{31}=0)$ because holes remain strongly localized in the WZ-ZB region. In addition, no pronounced $T_{SAW}$-periodic oscillations are observed in the data. This is consistent with the symmetric distribution of electrons immediately after photogeneration. When tuning the phase to $\phi=180°$, the band edge modulation at the time of photoexcitation is reversed. Now, holes redistribute slowly in both directions while electrons are created at a stable minimum in the conduction band. Albeit being less efficient, the sketched symmetric spatial redistribution of holes leads to a pronounced $T_{SAW}$-periodic modulation in both $t_{21}$ and $t_{31}$ because this transfer is symmetric in both directions. Thus, holes are trapped in twin defects on the left and in WZ segments on the right. As electrons are shuffled back and forth to the region with high rotational twin density (ZB/WZ) and the highly mixed WZ/ZB region, recombination via the two channels is modulated with $T_{SAW}$. At $\phi=90°$ and $\phi=270°$, the axial electric field amplitude is maximum and oriented parallel or antiparallel to the SAW wavevector, respectively. Thus, electrons are accelerated in opposite directions away from ($\phi=90°$) and towards ($\phi=270°$) the WZ/ZB region. These different initial conditions have pronounced impact on the $g^{(2)}_{31}$ data. For $\phi=90°$, when electrons are first transported away from the WZ/ZB segments, three-photon coincidences are found at $t_{21}=t_{31}=0$, analogous to the considerations made for $\phi=0$. However, for $\phi=90°$, the motion of the electrons occurs only in one direction. This different initial condition preserves the timing precision of the SAW-driven electron motion for $t_{21},t_{31}>0$. Therefore, the $T_{SAW}$-periodic modulation pattern is also preserved. For $\phi=270°$, electrons are accelerated towards the highly mixed WZ/ZB region on the right. There, they are reflected by the energy barriers in the conduction band at the interface between ZB- and WZ-GaAs. Thus, all indirect transitions are delayed by $T_{SAW}$ compared to the predominant ZB-emission at $t=0$. Thus, we expect that the maximum of three-photon coincidences occurs at $t_{21}=t_{31}=T_{SAW}$, which is precisely observed in the experimental data.

## 5 Conclusions and outlook

In summary, we implemented a fully-fledged combination of three-channel cross correlation and acousto-optoelectric spectroscopy (3CXC AOES). We used this method to probe coupled SAW-driven carrier dynamics in a single polytypic NW. Our method simultaneously records the SAW-induced modulations of all three emission channels. These data can be analyzed individually in a single channel analysis. Most importantly, as all channels are recorded simultaneously, second and third order correlations $g_{ij}^{(2)}(t_{ij})$ and $g_{1,2,3}^{(3)}(t_{21}, t_{31})$ can be faithfully determined from the data. We exemplify this advanced analysis for the three decay channels observed for our NW with sub-nanosecond resolution. These decay channels correspond to three distinct transitions in the type-II band edge modulation induced by crystal phase mixing in two different regions of the NW, a near pristine ZB region of the NW and a region consisting of extended WZ and ZB segments. Our analyses unambiguously resolve signatures of coupled carrier dynamics triggering an exchange between different regions within the NW. We observe correlated dynamics stemming from SAW-induced shuttling of electrons which recombine with holes trapped in the two regions. Moreover, we show that band edge modulation induced by the SAW at the time of photoexcitation programs the subsequent correlated dynamics. We point out that our method can be extended to other types of SAW-modulated processes. It is ideally suited to probe SAW-regulated carrier injection into and charge state control of optically active quantum dots [9,35–38]. Further semiconductor systems include acoustically-tunable coupled photonic structure [39,40], quantum dot-cavity coupling (light-matter interactions) [41] and quantum dot-optomechanical systems [42,43]. Moreover, it may find applications to probe SAW-driven carrier dynamics and SAW-induced charge transfer in 2D materials [44], van der Waals heterostructures [45,46], hybrid nanosystems [47–49] or SAW-generated acoustic lattices [50].

## Acknowledgements

This work was supported by the Deutsche Forschungsgemeinschaft (DFG) via the Research grants KR3790/6-1 and KO4005/6-1. M. M. S. and H. J. K. thank Achim Wixforth for his continuous support and valuable discussions.## References

[1] P. Delsing, A. N. Cleland, M. J. A. Schuetz, J. Knörzer, G. Giedke, J. I. Cirac, K. Srinivasan, M. Wu, K. C. Balram, C. Baüerle, T. Meunier, C. J. B. Ford, P. V. Santos, E. Cerda-Méndez, H. Wang, H. J. Krenner, E. D. S. Nysten, M. Weiß, G. R. Nash, L. Thevenard, C. Gourdon, P. Rovillain, M. Marangolo, J. Y. Duquesne, G. Fischerauer, W. Ruile, A. Reiner, B. Paschke, D. Denysenko, D. Volkmer, A. Wixforth, H. Bruus, M. Wiklund, J. Reboud, J. M. Cooper, Y. Q. Fu, M. S. Brugger, F. Rehfeldt, and C. Westerhausen, *The 2019 Surface Acoustic Waves Roadmap*, J. Phys. D. Appl. Phys. **52**, (2019).